\documentstyle[a4,12pt]{article}
\renewcommand{\baselinestretch}{1.225}
\def\thefootnote{\fnsymbol{footnote}}

\begin{document}
\parskip=5pt plus 1pt minus 1pt

\begin{flushright}
{\bf YITP-97-31} \\
{\bf DPNU-97-28}
\end{flushright}
\begin{flushright}
June 1997
\end{flushright}

\vspace{0.2cm}

\begin{center}
{\large\bf Flavor Symmetries and the Description of Flavor Mixing}
\end{center}

\vspace{0.3cm}

\begin{center}
{\bf Harald Fritzsch}  \footnote{Electronic address:
bm@hep.physik.uni-muenchen.de} \\
{\it Yukawa Institute for Theoretical Physics, Kyoto University,
Kyoto 606-01, Japan; ~ and} \\
{\it Sektion Physik, Universit$\ddot{a}$t M$\ddot{u}$nchen,
Theresienstra$\beta$e 37, D-80333 M$\ddot{u}$nchen, Germany}
\end{center}

\begin{center}
{\bf Zhi-zhong Xing} \footnote{Electronic address: xing@eken.phys.nagoya-u.ac.jp} \\
{\it Department of Physics, Nagoya University, Chikusa-ku, Nagoya 464-01, Japan}
\end{center}

\vspace{1.8cm}

\begin{abstract}
It is shown that the hierarchical structure of the quark mass terms in the
standard model suggests a new description of the flavor mixing. The
latter is primarily  a heavy quark mixing involving the $t$ and $b$ quarks,
followed by a mixing exclusively in the u-channel or the d-channel. The  complex
phase  describing $CP$  violation arises only in the light quark sector. The
Cabibbo angle is not a basic parameter, but  results as a superposition of
both the u-channel and d-channel mixing terms. The new description has
a number of significant advantages in comparison with all descriptions
previously used. It is suggested that the new description be used 
in all future discussions of flavor physics and $CP$ violation. 
\end{abstract}

\newpage

A deeper understanding of the phenomenon of flavor mixing observed in the
charged-current--type weak interactions remains one of the major challenges
in particle physics at present. In the standard electroweak theory it is
described by a 3 $\times $ 3 unitary flavor mixing matrix 
\cite{KM,Standard}, which
can be expressed in terms of four parameters, usually taken to be as three rotation
angles and a phase angle. Within the standard model there is no way to obtain
any further information about these parameters. Any attempt to do so would
imply physics inputs which go beyond the standard electroweak theory.

In the standard model the mixing of quark flavors arises after the diagonalization of
the up- and down-type mass matrices. Both mass matrices cannot be
diagonalized by unitary transformations which commute with the charged weak
generators. The result of this diagonalization-mismatch, whose dynamical
origin is unknown, is the flavor mixing. However, it is implied that the
mechanism which is responsible for the generation of quark masses is at
the same time responsible for the mixing of flavors, i.e., any change of the
eigenvalues of quark masses would in general also lead to a change of the
flavor mixing parameters. In many models based on flavor symmetries which go
beyond the standard electroweak theory, the flavor mixing parameters are
indeed functions of the mass eigenvalues (for early works, see Refs. 
\cite{Fritzsch77,Weinberg77,Fritzsch79}).

Both the observed mass spectrum of quarks and the observed values of 
flavor mixing parameters exhibit a striking hierarchical structure. This
hierarchical structure can be understood in a natural way as the result of
a specific pattern of chiral symmetries whose breaking would cause the
hierarchical tower of masses to appear step by step 
\cite{Fritzsch87a,Fritzsch87b,Hall93a}. Such a
chiral evolution of the quark mass matrices leads, as argued in particular in
Ref. \cite{Fritzsch87b}, 
to a rather specific way to describe the flavor mixing; in the
limit $m_u = m_d = 0 $ the flavor mixing is merely a rotation between the
second and third families, described by one rotation angle. While the
original representation of the flavor mixing, introduced in Ref. \cite{KM}, is
not natural in the sense of a chiral evolution of the mass matrices, the
standard representation \cite{Standard}
does respect it. Nevertheless, there still remain several possibilities to
describe the flavor mixing. In particular the complex phase which describes
$CP$ violation can appear in a number of different ways, and in general the
relation between this phase and the phases of quark mass terms is rather
complicated.

In this paper we should like to point out that there exists a
parametrization of the flavor mixing which is unique in the sense that 
it incorporates the chiral evolution of the mass matrices in a 
natural way, and that the phase in the flavor mixing matrix and the
phases in the mass terms are related in a very simple
way. Hence the phase entering the flavor mixing matrix could have a
deeper physical meaning.
In addition, it turns out that the mixing parameters introduced
here can be described by quantities which are easily measurable in 
$B$-meson decays. Related descriptions of the flavor mixing have
appeared in approximate forms before
\cite{Fritzsch79,Dimopoulos92,Hall93b}. The arguments presented below 
are, in our view, strong enough in order to reconsider
the way how the mixing of quark flavors should be described and
parametrized.

In this paper we take the point of view that the quark mass
eigenvalues are dynamical entities, and one could change their values
in order to study certain symmetry limits, as it is done in QCD. In
the standard electroweak model, in which the quark mass matrices are
given by the coupling of a scalar field to various quark fields,
this can certainly be done by adjusting the related coupling
constants. Whether it is possible in reality is an open question.

It is well-known that the quark mass matrices can always be made hermitian
by a suitable transformation of the right-handed fields. Without loss
of generality, we shall
suppose in this paper that the quark mass matrices are hermitian.

In the limit where the masses of the $u$ and $d$ quarks are set to
zero, the quark mass matrix $\tilde{M}$ (for both charge $+2/3$ and
charge $-1/3$ sectors) can be arranged such that its elements 
$\tilde{M}_{i1}$ and $\tilde{M}_{1i}$ ($i=1,2,3$) are all zero 
\cite{Fritzsch87a,Fritzsch87b}. Thus the quark
mass matrices have the form
\begin{equation}
\tilde{M} \; =\; \left ( \matrix{
0	& 0	& 0 \cr
0	& \tilde{C}	& \tilde{B} \cr
0	& \tilde{B}^*	& \tilde{A} \cr} \right ) \; .
\end{equation}
The observed mass hierarchy is incorporated into this structure by
denoting the entry which is of the order of the $t$-quark or 
$b$-quark mass by $\tilde{A}$, with $\tilde{A}\gg \tilde{C},
|\tilde{B}|$. It can easily be seen (see, e.g., Ref. \cite{Lehmann96}) that
the complex phases in the mass matrices (1) can be
rotated away by subjecting both $\tilde{M}_{\rm u}$ and $\tilde{M}_{\rm d}$ to the
same unitary transformation. Thus we shall take $\tilde{B}$ to be
real for both up- and down-quark sectors. As expected, $CP$ violation
cannot arise at this stage. The diagonalization of the mass matrices
leads to a mixing between the second and third families, described by an
angle $\tilde{\theta}$. The flavor mixing matrix is
then given by
\begin{equation}
\tilde{V} \; =\; \left ( \matrix{
1	& 0	& 0 \cr
0	& \tilde{c}	& \tilde{s} \cr
0	& -\tilde{s}	& \tilde{c} \cr } \right ) \; ,
\end{equation}
where $\tilde{s} \equiv \sin \tilde{\theta}$ and $\tilde{c} \equiv
\cos \tilde{\theta}$. In view of the fact that the limit $m_u = m_d
=0$ is not far from reality, the angle $\tilde{\theta}$ is essentially 
given by the observed value of $|V_{cb}|$ ($=0.039 \pm 0.002$ \cite{Neubert96,Forty97});
i.e., $\tilde{\theta} = 2.24^{\circ} \pm 0.12^{\circ}$.

At the next and final stage of the chiral evolution of the mass matrices,
the masses of the $u$ and $d$ quarks are introduced.
The hermitian mass matrices have in general the
form:
\begin{equation}
M \; =\; \left ( \matrix{
E	& D	& F \cr
D^*	& C	& B \cr
F^*	& B^*	& A \cr } \right ) \; 
\end{equation}
with $A\gg C, |B| \gg E, |D|, |F|$. By a common unitary transformation of 
the up- and down-type quark fields, one can always arrange the mass
matrices $M_{\rm u}$ and $M_{\rm d}$ in such a way that $F_{\rm u} =
F_{\rm d} =0$; i.e.,
\begin{equation}
M \; =\; \left ( \matrix{
E	& D	& 0 \cr
D^*	& C	& B \cr
0	& B^*	& A \cr } \right ) \; .
\end{equation}
This can easily be seen as follows. If phases are neglected, the two
symmetric mass matrices $M_{\rm u}$ and $M_{\rm d}$ can be transformed 
by an orthogonal transformation matrix $O$, which can be described by
three angles, such that they assume the form (4). The condition
$F_{\rm u} =F_{\rm d} =0$ gives two constraints for the three angles of 
$O$. If complex phases are allowed in $M_{\rm u}$ and $M_{\rm d}$, the 
condition $F_{\rm u} =F_{\rm u}^* = F_{\rm d} =F_{\rm d}^* =0$ imposes
four constraints, which can also be fulfilled, if $M_{\rm u}$ and
$M_{\rm d}$ are subjected to a common unitary transformation matrix $U$. The
latter depends on nine parameters. Three of them are not suitable for
our purpose, since they are just diagonal phases; but the remaining
six can be chosen such that the vanishing of $F_{\rm u}$ and $F_{\rm
d}$ results.

The basis in which the mass matrices take the form (4) is a basis in
the space of quark flavors, which in our view is of special
interest. It is a basis in which the mass matrices exhibit two
texture zeros, for both up- and down-type quark sectors. 
These, however, do not imply special relations among
mass eigenvalues and flavor mixing parameters (as pointed out
above). In this basis the mixing is of the ``nearest neighbour'' form, 
since the (1,3) and (3,1) elements of $M_{\rm u}$ and $M_{\rm d}$
vanish; no direct mixing between the heavy $t$ (or $b$) quark and the
light $u$ (or $d$) quark is present (see also Ref. \cite{Branco89}).
In certain models (see, e.g., Refs. \cite{Fritzsch79,Dimopoulos92}),
this basis is indeed of particular interest, but we shall proceed without 
relying on a special texture models for the mass matrices.

A mass matrix of the type (4) can in the absence of complex phases be
diagonalized by a rotation matrix, described by two angles only.
At first the off-diagonal element 
$B$ is rotated away by a rotation between the second and third 
families (angle $\theta_{23}$); at the second step the element $D$ is rotated away by a
transformation of the first and second families (angle $\theta_{12}$). No rotation between
the first and third families is required. 
The rotation matrix for this sequence takes the form
\begin{eqnarray}
R \; =\; R_{12} R_{23} & = & \left ( \matrix{
c_{12} 	& s_{12}	& 0 \cr
-s_{12}	& c_{12}	& 0 \cr
0	& 0	& 1 \cr } \right )  \left ( \matrix{
1	& 0	& 0 \cr
0	& c_{23}	& s_{23} \cr
0	& -s_{23}	& c_{23} \cr } \right ) \; 
\nonumber \\ \nonumber \\
& = & \left ( \matrix{
c_{12}	& s_{12} c_{23}	& s_{12} s_{23} \cr 
-s_{12}	& c_{12} c_{23} & c_{12} s_{23} \cr
0	& -s_{23}	& c_{23} \cr } \right ) \; ,
\end{eqnarray}
where $c_{12} \equiv \cos \theta_{12}$, $s_{12} \equiv \sin
\theta_{12}$, etc.
The flavor mixing matrix $V$ is the product of two such matrices, one
describing the rotation among the up-type quarks, and the other describing
the rotation among the down-type quarks:
\begin{equation}
V \; =\; R^{\rm u}_{12} R^{\rm u}_{23} ( R^{\rm d}_{23} )^{-1} ( R^{\rm d}_{12} )^{-1} \; .
\end{equation}
The product $R^{\rm u}_{23} (R^{\rm d}_{23} )^{-1}$ can be written as
a rotation matrix described by a single angle $\theta$. In the limit
$m_u = m_d =0$, this is just the angle $\tilde{\theta}$ encountered
in Eq. (2). The angle which describes the $R^{\rm u}_{12}$ rotation shall
be denoted by $\theta_{\rm u}$; and the corresponding angle for the
$R^{\rm d}_{12}$ rotation by $\theta_{\rm d}$. Thus in the absence of
$CP$-violating phases the flavor mixing matrix takes the following 
specific form:
\begin{eqnarray}
V & = & \left ( \matrix{
c_{\rm u}	& s_{\rm u}	& 0 \cr
-s_{\rm u}	& c_{\rm u}	& 0 \cr
0	& 0	& 1 \cr } \right )  \left ( \matrix{
1	& 0	& 0 \cr
0	& c	& s \cr
0	& -s	& c \cr } \right )  \left ( \matrix{
c_{\rm d}	& -s_{\rm d}	& 0 \cr
s_{\rm d}	& c_{\rm d}	& 0 \cr
0	& 0	& 1 \cr } \right )  \nonumber \\ \nonumber \\
& = & \left ( \matrix{
s_{\rm u} s_{\rm d} c + c_{\rm u} c_{\rm d} 	& 
s_{\rm u} c_{\rm d} c - c_{\rm u} s_{\rm d} 	& s_{\rm u} s \cr
c_{\rm u} s_{\rm d} c - s_{\rm u} c_{\rm d} 	& 
c_{\rm u} c_{\rm d} c + s_{\rm u} s_{\rm d} 	& c_{\rm u} s \cr
-s_{\rm d} s	& -c_{\rm d} s	& c \cr } \right ) \; ,
\end{eqnarray}
where $c_{\rm u} \equiv \cos\theta_{\rm u}$, $s_{\rm u} \equiv
\sin\theta_{\rm u}$, etc.

We proceed by including the phase parameters of the quark mass
matrices in Eq. (4). Each mass matrix has in general two complex 
phases. These phases can be dealt with in a similar way as described
in Refs. \cite{Fritzsch77,Fritzsch79}. It can easily be seen that, 
by suitable rephasing of the quark fields,
the flavor mixing matrix can finally be written in terms of only a
single phase $\varphi$ as follows:
\begin{eqnarray}
V & = & \left ( \matrix{
c_{\rm u}	& s_{\rm u}	& 0 \cr
-s_{\rm u}	& c_{\rm u}	& 0 \cr
0	& 0	& 1 \cr } \right )  \left ( \matrix{
e^{-{\rm i}\varphi}	& 0	& 0 \cr
0	& c	& s \cr
0	& -s	& c \cr } \right )  \left ( \matrix{
c_{\rm d}	& -s_{\rm d}	& 0 \cr
s_{\rm d}	& c_{\rm d}	& 0 \cr
0	& 0	& 1 \cr } \right )  \nonumber \\ \nonumber \\
& = & \left ( \matrix{
s_{\rm u} s_{\rm d} c + c_{\rm u} c_{\rm d} e^{-{\rm i}\varphi} &
s_{\rm u} c_{\rm d} c - c_{\rm u} s_{\rm d} e^{-{\rm i}\varphi} &
s_{\rm u} s \cr
c_{\rm u} s_{\rm d} c - s_{\rm u} c_{\rm d} e^{-{\rm i}\varphi} &
c_{\rm u} c_{\rm d} c + s_{\rm u} s_{\rm d} e^{-{\rm i}\varphi}   &
c_{\rm u} s \cr
- s_{\rm d} s	& - c_{\rm d} s	& c \cr } \right ) \; .
\end{eqnarray}
Note that the three angles $\theta_{\rm u}$, $\theta_{\rm d}$ and
$\theta$ in Eq. (8) can all be arranged to lie in the first quadrant
through a suitable redefinition of quark field phases. Consequently
all $s_{\rm u}$, $s_{\rm d}$, $s$ and $c_{\rm u}$, $c_{\rm d}$, $c$
are positive. The phase $\varphi$ can in general take values from 0
to $2\pi$; and $CP$ violation is present in weak interactions
if $\varphi \neq 0, \pi$ and $2\pi$.

This particular representation of the flavor mixing matrix is the main result of
this paper. In comparison with all other parametrizations discussed
previously \cite{KM,Standard}, it
has a number of interesting features which in our view make it very
attractive and provide strong arguments for its use in future
discussions of flavor mixing phenomena, in particular, those in
$B$-meson physics. We shall discuss them below.

a) The flavor mixing matrix $V$ in Eq. (8) follows directly from the
chiral expansion of the mass
matrices. Thus it naturally takes into account the hierarchical structure of the 
quark mass spectrum.

b) The complex phase describing $CP$ violation ($\varphi$) appears only in the
(1,1), (1,2), (2,1) and (2,2) elements of $V$, i.e., 
in the elements involving only the quarks of the first and second
families. This is a natural description of $CP$ violation since in our 
hierarchical approach $CP$ violation is not directly linked to the third family, but
rather to the first and second ones, and in particular to the mass terms of the
$u$ and $d$ quarks.

It is instructive to consider the special case $s_{\rm u} = s_{\rm d}
= s = 0$. Then the flavor mixing matrix $V$ takes the form
\begin{equation}
V \; = \; \left ( \matrix{
e^{-{\rm i}\varphi}	& 0	& 0 \cr
0	& 1	& 0 \cr
0	& 0	& 1 \cr} \right ) \; .
\end{equation}
This matrix describes a phase change in the weak transition between
$u$ and $d$, while no phase change is present in the
transitions between $c$ and $s$ as well as $t$ and $b$.
Of course, this effect can be absorbed in a phase change of the $u$-
and $d$-quark fields, and no $CP$ violation is present. Once the
angles $\theta_{\rm u}$, $\theta_{\rm d}$ and $\theta$ are introduced, 
however, $CP$ violation arises. It is due to a phase change in the weak
transition between $u^{\prime}$ and $d^{\prime}$, where $u^{\prime}$
and $d^{\prime}$ are the rotated quark fields, obtained by applying
the corresponding rotation matrices given in Eq. (8) to the 
quark mass eigenstates ($u^{\prime}$: mainly $u$, small admixture of
$c$; $d^{\prime}$: mainly $d$, small admixture of $s$).

Since the mixing matrix elements involving $t$ or $b$ quark are real
in the representation (8), one can find that the phase parameter of
$B^0_q$-$\bar{B}^0_q$ mixing ($q=d$ or $s$), dominated by the
box-diagram contributions in the standard model \cite{PDG96}, is essentially unity:
\begin{equation}
\left ( \frac{q}{p} \right )_{B_q} = \;
\frac{V^*_{tb}V_{tq}}{V_{tb}V^*_{tq}} \; = \; 1 \; .
\end{equation}
In most of other parametrizations of the flavor mixing matrix,
however, the two rephasing-variant quantities 
$(q/p)^{~}_{B_d}$ and $(q/p)^{~}_{B_s}$ take different (maybe complex) values.

c) The dynamics of flavor mixing can easily be interpreted by
considering certain limiting cases in Eq. (8). In the limit $\theta
\rightarrow 0$ (i.e., $s \rightarrow 0$ and $c\rightarrow 1$), the
flavor mixing is, of course, just a mixing between the first and
second families, described by only one mixing angle (the Cabibbo angle 
$\theta_{\rm C}$ \cite{Cabibbo63}).  
It is a special and essential feature of the representation (8) that the Cabibbo
angle is {\it not} a basic angle, used in the parametrization. 
The matrix element $V_{us}$ (or $V_{cd}$) is
indeed a superposition of two terms including a phase. This feature
arises naturally in our hierarchical approach, but it is not new. In
many models of specific textures of mass matrices, it is indeed the
case that the Cabibbo-type transition $V_{us}$ (or $V_{cd}$) 
is a superposition of several
terms. At first, it was obtained by one of the authors 
in the discussion of the two-family mixing \cite{Fritzsch77}.

In the limit $\theta =0$ considered here, one has $|V_{us}| = |V_{cd}|
= \sin\theta_{\rm C} \equiv s^{~}_{\rm C}$ and
\begin{equation}
s^{~}_{\rm C} \; =\; \left | s_{\rm u} c_{\rm d} ~ - ~ c_{\rm u} s_{\rm d}
e^{-{\rm i}\varphi} \right | \; .
\end{equation}
This relation describes a triangle in the complex plane, as
illustrated in Fig. 1, which we
shall denote as the ``Cabibbo triangle''. 
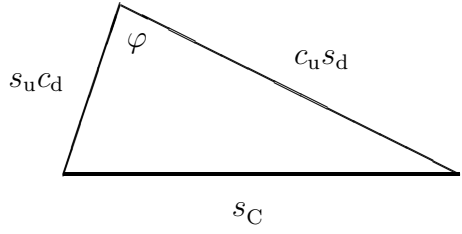
\begin{figure}[t]
\begin{picture}(400,160)(-90,210)
\put(80,300){\line(1,0){150}}
\put(80,300.5){\line(1,0){150}}
\put(150,285.5){\makebox(0,0){$s^{~}_{\rm C}$}}
\put(80,300){\line(1,3){21.5}}
\put(80,300.5){\line(1,3){21.5}}
\put(80,299.5){\line(1,3){21.5}}
\put(70,335){\makebox(0,0){$s_{\rm u} c_{\rm d}$}}
\put(230,300){\line(-2,1){128}}
\put(230,300.5){\line(-2,1){128}}
\put(178,343.5){\makebox(0,0){$c_{\rm u} s_{\rm d}$}}
\put(108,350){\makebox(0,0){$\varphi$}}
\end{picture}
\vspace{-2.5cm}
\caption{The Cabibbo triangle in the complex plane.}
\end{figure}
This triangle is a feature of
the mixing of the first two families (see also
Ref. \cite{Fritzsch77}). Explicitly one has (for $s=0$):
\begin{equation}
\tan\theta_{\rm C} \; =\; \sqrt{\frac{\tan^2\theta_{\rm u} +
\tan^2\theta_{\rm d} - 2 \tan\theta_{\rm u} \tan\theta_{\rm d}
\cos\varphi}
{1 + \tan^2\theta_{\rm u} \tan^2\theta_{\rm d} + 2 \tan\theta_{\rm u}
\tan\theta_{\rm d} \cos\varphi}} \; .
\end{equation}
Certainly the flavor mixing matrix $V$ cannot accommodate $CP$ violation in this
limit. However, the existence of $\varphi$ seems necessary in order
to make Eq. (12) compatible with current data, as one can see below.

d) The three mixing angles $\theta$, $\theta_{\rm u}$ and 
$\theta_{\rm d}$ have a precise physical meaning. The angle $\theta$
describes the mixing between the second and third families, which is
generated by the off-diagonal terms $B_{\rm u}$ and $B_{\rm d}$ in the 
up and down mass matrices of Eq. (4). 
We shall refer to this mixing involving $t$ and $b$ as the ``heavy
quark mixing''.
The angle $\theta_{\rm u}$,
however, solely describes the $u$-$c$ mixing, corresponding to the $D_{\rm
u}$ term in $M_{\rm u}$. We shall denote this as the ``u-channel mixing''.
The angle $\theta_{\rm d}$ solely describes 
the $d$-$s$ mixing, corresponding to the $D_{\rm d}$ term in $M_{\rm
d}$; this will be denoted as the ``d-channel mixing''. 
Thus there exists an asymmetry between the mixing of the first and
second families and that of the second and third families,
which in our view reflects interesting details of the underlying dynamics of
flavor mixing. 
The heavy quark mixing is a combined effect, involving both charge
$+2/3$ and charge $-1/3$ quarks, while the u- or d-channel mixing
(described by the angle $\theta_{\rm u}$ or $\theta_{\rm d}$) proceeds 
solely in the charge $+2/3$ or charge $-1/3$ sector. Therefore an
experimental determination of these two angles would allow to draw
interesting conclusions about the amount and perhaps the underlying
pattern of the u- or d-channel mixing.

e) The three angles $\theta$, $\theta_{\rm u}$ and $\theta_{\rm d}$
are related in a very simple way to observable quantities of $B$-meson 
physics. 
For example, $\theta$ is related to 
the rate of the semileptonic decay $B\rightarrow D^*l\nu^{~}_l$; 
$\theta_{\rm u}$ is associated with the ratio of the decay rate of
$B\rightarrow (\pi, \rho) l \nu^{~}_l$ to that of $B\rightarrow 
D^* l\nu^{~}_l$; and $\theta_{\rm d}$ can be determined from the ratio of
the mass difference between two $B_d$ mass eigenstates to that between
two $B_s$ mass eigenstates. From Eq. (8) we find the following exact
relations:
\begin{equation}
\sin \theta \; = \; |V_{cb}| \sqrt{ 1 + \left |\frac{V_{ub}}{V_{cb}}
\right |^2} \; ,
\end{equation}
and
\begin{eqnarray}
\tan\theta_{\rm u} & = & \left | \frac{V_{ub}}{V_{cb}} \right | \; ,
\nonumber \\
\tan\theta_{\rm d} & = & \left | \frac{V_{td}}{V_{ts}} \right | \; .
\end{eqnarray}
These simple results make the parametrization (8) uniquely favorable 
for the study of $B$-meson physics.

By use of current data on $|V_{ub}|$ and $|V_{cb}|$, i.e., $|V_{cb}| = 
0.039 \pm 0.002$ \cite{Neubert96,Forty97} and $|V_{ub}/V_{cb}| =0.08 \pm 0.02$ 
\cite{PDG96}, we obtain $\theta_{\rm u} = 4.57^{\circ} \pm
1.14^{\circ}$ and $\theta = 2.25^{\circ} \pm 0.12^{\circ}$. Taking
$|V_{td}| = (8.6 \pm 2.1) \times 10^{-3}$ \cite{Forty97},
which was obtained from the analysis of current data on
$B^0_d$-$\bar{B}^0_d$ mixing,
we get $|V_{td}/V_{ts}| = 0.22 \pm 0.07$, i.e., $\theta_{\rm d} = 12.7^{\circ} 
\pm 3.8^{\circ}$.
Both the heavy quark mixing angle $\theta$ and the u-channel mixing
angle $\theta_{\rm u}$ are relatively small. The smallness of $\theta$ 
implies that Eqs. (11) and (12) are valid to a high degree of
precision (of order $1-c \approx 0.001$).

f) It is instructive to consider the limiting case $\theta_{\rm u}
\rightarrow 0$ or $\theta_{\rm d} \rightarrow 0$, which can be
achieved by setting $D_{\rm u} \rightarrow 0$ or $D_{\rm d}
\rightarrow 0$ in Eq. (4). In the absence of the u-channel mixing
($\theta_{\rm u} =0$), one has $V_{ub}=0$ \cite{Fritzsch85}; in the absence of the
d-channel mixing, $V_{td} =0$ appears. In both cases $CP$ violation is 
absent. In reality, we have $\theta_{\rm d}/\theta_{\rm u} \sim 2
\cdot\cdot\cdot 3$, i.e., the u-channel mixing is significantly
smaller than the d-channel mixing.

In the absence of the u-channel mixing ($\theta_{\rm u}=0$), one finds 
\begin{equation}
\left | \frac{V_{us}}{V_{ud}} \right | \; =\; \left |
\frac{V_{cd}}{V_{cs}} \right | \; =\; \left | \frac{V_{td}}{V_{ts}}
\right | \; =\; \tan\theta_{\rm d} \; ;
\end{equation}
and in the absence of the d-channel mixing ($\theta_{\rm d} = 0$), one arrives at
\begin{equation}
\left | \frac{V_{us}}{V_{cs}} \right | \; =\; \left |
\frac{V_{cd}}{V_{ud}} \right | \; =\; \left | \frac{V_{ub}}{V_{cb}}
\right | \; =\; \tan\theta_{\rm u} \; .
\end{equation}
Certainly the last relation does not hold well, since the experimental 
value of $|V_{ub}/V_{cb}|$ is only about $1/3$ of that of 
$|V_{us}/V_{cs}|$. Of course, this is due to the fact
that the d-channel mixing angle $\theta_{\rm d}$ is relatively large. 
In contrast, the relation (15) is consistent with current
data. If it were exact, i.e.,  $\theta_{\rm u} = 0$, then $\tan\theta_{\rm 
d}$ would be determined by $|V_{us}/V_{ud}|$,
which has been precisely measured. Using the experimental values 
$|V_{us}| = 0.2205 \pm 0.0018$ and $|V_{ud}| = 0.9736 \pm 0.0010$
\cite{PDG96}, one would find $\theta_{\rm d} = 12.76^{\circ} \pm
0.12^{\circ}$ on the basis of Eq. (15). However, since $\theta_{\rm u}$ does not
vanish exactly, the actual error for $\theta_{\rm d}$ obtained from Eq. (15) 
should be as large as that given above:
$\theta_{\rm d} = 12.7^{\circ} \pm 3.8^{\circ}$. 
It is interesting, nevertheless, that the central
values of $\theta_{\rm d}$ obtained from the relations (14) and (15)
are essentially identical.

g) According to Eq. (8), as well as Eq. (11), the phase $\varphi$ is
a phase difference between the contributions to $V_{us}$ (or $V_{cd}$) 
from the u-channel mixing and the d-channel mixing. Therefore
$\varphi$ is given by the relative phase of $D_{\rm d}$ and $D_{\rm
u}$ in the quark mass matrices (4), if the phases of $B_{\rm u}$ and
$B_{\rm d}$ are absent or negligible.

The phase $\varphi$ is not likely to be $0^{\circ}$ or $180^{\circ}$, according
to the experimental values given above, even though the measurement of 
$CP$ violation in $K^0$-$\bar{K}^0$ mixing \cite{PDG96} is not taken
into account. For $\varphi =0^{\circ}$, one
finds $\tan\theta_{\rm C} = 0.14 \pm 0.08$; and for $\varphi =
180^{\circ}$, one gets $\tan\theta_{\rm C} = 0.30 \pm 0.08$. Both
cases are barely consistent with the value of $\tan\theta_{\rm
C}$ obtained from experiments ($\tan\theta_{\rm C} \approx
|V_{us}/V_{ud}| \approx 0.226$).

h) The $CP$-violating phase $\varphi$ in the flavor mixing matrix $V$ can be
determined from $|V_{us}|$ ($= 0.2205 \pm 0.0018$ \cite{PDG96})
through the following formula, obtained easily from Eq. (8):
\begin{equation}
\varphi \; =\; \arccos \left ( \frac{s^2_{\rm u} c^2_{\rm d} c^2 +
c^2_{\rm u} s^2_{\rm d} - |V_{us}|^2}{2 s_{\rm u} c_{\rm u} s_{\rm d}
c_{\rm d} c} \right ) \; .
\end{equation}
The two-fold ambiguity associated with the value of $\varphi$, coming
from $\cos\varphi = \cos (2\pi - \varphi)$, is removed if one
takes $\sin\varphi >0$ into account (this is required by current data on $CP$ violation in
$K^0$-$\bar{K}^0$ mixing (i.e., $\epsilon^{~}_K$) \cite{PDG96}).
More precise measurements of the angles $\theta_{\rm u}$ and $\theta_{\rm
d}$ in the forthcoming experiments of $B$ physics will remarkably reduce the 
uncertainty of $\varphi$ to be determined from Eq. (17). This approach is of 
course complementary to the direct determination of $\varphi$ from
$CP$ asymmetries in some weak $B$-meson decays into hadronic $CP$
eigenstates \cite{Sanda80}.

\begin{figure}
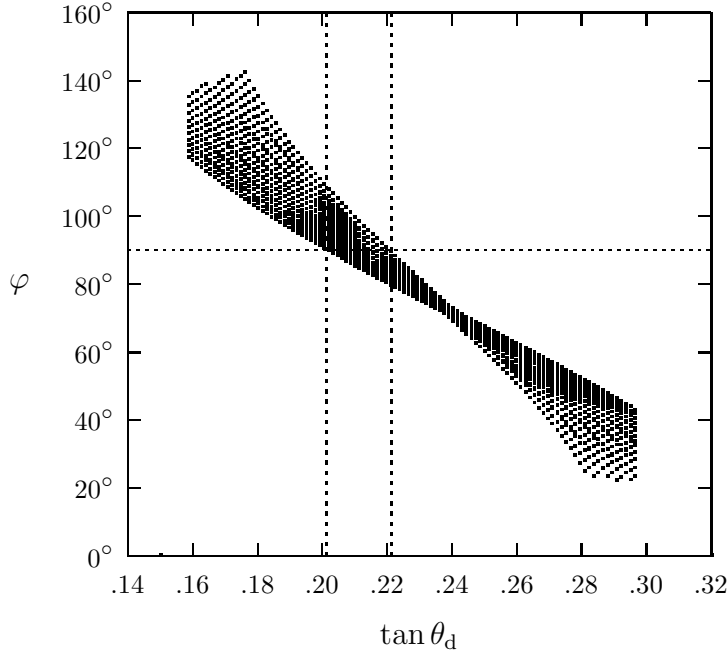

\setlength{\unitlength}{0.240900pt}
\ifx\plotpoint\undefined\newsavebox{\plotpoint}\fi
\sbox{\plotpoint}{\rule[-0.200pt]{0.400pt}{0.400pt}}%

\vspace{0.4cm}
\caption{The region of the $CP$-violating phase $\varphi$ allowed by 
current data.}
\end{figure}

For illustration, we plot the allowed region of $\varphi$ as a function of
$\tan\theta_{\rm d}$ in Fig. 2, where the values of $|V_{us}|$, $\theta$,
$\theta_{\rm u}$ and $\theta_{\rm d}$ are taken to vary independently
within their corresponding errors. The constraint from the
experimental result of $\epsilon^{~}_K$ on $\varphi$ is also included
(see, e.g., Ref. \cite{Buras96} for relevant formulas of $\epsilon^{~}_K$).
One can find that the value of $\varphi$ is most likely in the range $40^{\circ}$
to $120^{\circ}$; indeed the central values of the four
inputs lead to $\varphi \approx 81^{\circ}$. 
Note that $\varphi$ is essentially independent of the angle $\theta$,
due to the tiny observed value of the latter. 
Once $\tan\theta_{\rm d}$ is precisely measured, we shall be able to fix the
magnitude of $\varphi$ to a satisfactory degree of accuracy.

i) It is well-known that $CP$ violation in the flavor mixing matrix
$V$ can be rephasing-invariantly described by a universal quantity
$\cal J$ \cite{Jarlskog85}:
\begin{equation}
{\rm Im} \left ( V_{il} V_{jm} V^*_{im} V^*_{jl} \right ) \; =\;
{\cal J} \sum_{k,n=1}^{3} \left ( \epsilon^{~}_{ijk} \epsilon^{~}_{lmn} \right ) \; .
\end{equation}
In the parametrization (8), $\cal J$ reads
\begin{equation}
{\cal J} \; =\; s_{\rm u} c_{\rm u} s_{\rm d} c_{\rm d} s^2 c \sin\varphi \; .
\end{equation}
Obviously $\varphi=90^{\circ}$ leads to the maximal value of $\cal J$.

Indeed $\varphi =90^{\circ}$, a particularly interesting case for $CP$ 
violation, is quite consistent with
current data. One can see from Fig. 2 that this possibility exists
if $0.202 \leq \tan\theta_{\rm d} 
\leq 0.222$, or $11.4^{\circ} \leq \theta_{\rm d} \leq 12.5^{\circ}$.
In this case the mixing term
$D_{\rm d}$ in Eq. (4) can be taken to be real, and the term $D_{\rm 
u}$ to be imaginary, if ${\rm Im}(B_{\rm u}) = {\rm Im} (B_{\rm d})
=0$ is assumed (see also Refs. \cite{FritzschXing95,Lehmann96}). 
Since in our description of the flavor mixing the
complex phase $\varphi$ is related in a simple way to the phases of
the quark mass terms, the case $\varphi = 90^{\circ}$ is especially
interesting. It can hardly be an accident, and this case should be
studied further. The possibility that the phase $\varphi$ describing
$CP$ violation in the standard model is given by the algebraic number
$\pi/2$ should be taken seriously. It may provide a useful clue
towards a deeper understanding of the origin of $CP$ violation
and of the dynamical origin of the fermion masses.

In Ref. \cite{FritzschXing95} the case $\varphi =90^{\circ}$ has been
denoted as ``maximal'' $CP$ violation. It implies in our framework 
that in the complex
plane the u-channel and d-channel mixings are perpendicular to each
other. In this special case (as well as $\theta\rightarrow 0$), we have 
\begin{equation}
\tan^2\theta_{\rm C} \; =\; \frac{\tan^2\theta_{\rm u} ~ + ~
\tan^2\theta_{\rm d}}{1 ~ + ~ \tan^2\theta_{\rm u} \tan^2\theta_{\rm
d}} \; .
\end{equation}
To a good approximation (with the relative error $\sim 2\%$), 
one finds $s^2_{\rm C} \approx s^2_{\rm u} + s^2_{\rm d}$. 

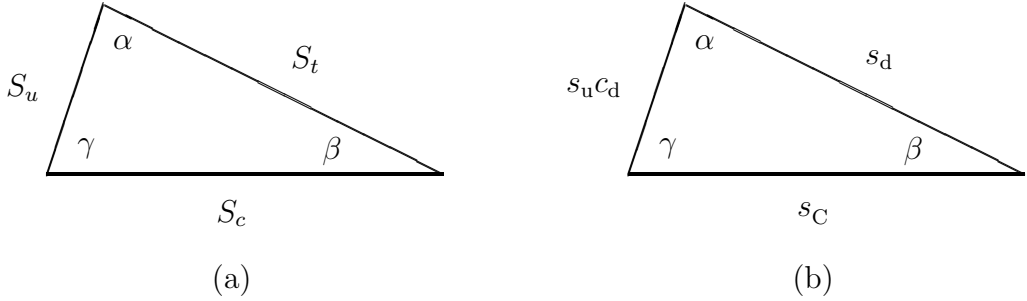
\begin{figure}[t]
\begin{picture}(400,160)(10,210)
\put(80,300){\line(1,0){150}}
\put(80,300.5){\line(1,0){150}}
\put(150,285.5){\makebox(0,0){$S_c$}}
\put(80,300){\line(1,3){21.5}}
\put(80,300.5){\line(1,3){21.5}}
\put(80,299.5){\line(1,3){21.5}}
\put(71,333){\makebox(0,0){$S_u$}}
\put(230,300){\line(-2,1){128}}
\put(230,300.5){\line(-2,1){128}}
\put(178,343.5){\makebox(0,0){$S_t$}}
\put(95,310){\makebox(0,0){$\gamma$}}
\put(188,309){\makebox(0,0){$\beta$}}
\put(109,350){\makebox(0,0){$\alpha$}}
\put(150,260){\makebox(0,0){(a)}}
\put(300,300){\line(1,0){150}}
\put(300,300.5){\line(1,0){150}}
\put(370,285.5){\makebox(0,0){$s^{~}_{\rm C}$}}
\put(300,300){\line(1,3){21.5}}
\put(300,300.5){\line(1,3){21.5}}
\put(300,299.5){\line(1,3){21.5}}
\put(287,333){\makebox(0,0){$s_{\rm u} c_{\rm d}$}}
\put(450,300){\line(-2,1){128}}
\put(450,300.5){\line(-2,1){128}}
\put(395,343.5){\makebox(0,0){$s_{\rm d}$}}
\put(315,310){\makebox(0,0){$\gamma$}}
\put(408,309){\makebox(0,0){$\beta$}}
\put(329,350){\makebox(0,0){$\alpha$}}
\put(370,260){\makebox(0,0){(b)}}
\end{picture}
\vspace{-1.5cm}
\caption{The unitarity triangle (a) and its rescaled counterpart (b)
in the complex plane.}
\end{figure}

j) At future $B$-meson factories, the study of $CP$ violation will
concentrate on measurements of the unitarity triangle 
\begin{equation}
S_u ~ + ~ S_c ~ + ~ S_t \; = \; 0 \; ,
\end{equation}
where $S_i \equiv V_{id} V^*_{ib}$ in the complex
plane (see Fig. 3(a) for illustration). 
The inner angles of this triangle are denoted as \cite{PDG96}
\footnote{An alternative notation for three angles of the unitarity
triangle is $(\phi_1, ~ \phi_2, ~ \phi_3)$, equivalent to $(\beta, ~
\alpha, ~ \gamma)$.}
\begin{eqnarray}
\alpha & \equiv & \arg (- S_t S^*_u ) \; , \nonumber \\
\beta  & \equiv & \arg (- S_c S^*_t ) \; , \nonumber \\
\gamma & \equiv & \arg (- S_u S^*_c ) \; .
\end{eqnarray}
Obviously $\alpha + \beta + \gamma = \pi$ is a trivial
consequence of the above definition \cite{Xing96}. In terms of the parameters
$\theta$, $\theta_{\rm u}$, $\theta_{\rm d}$ and $\varphi$, we obtain
\begin{eqnarray}
\sin (2\alpha) & = & \frac{2 c_{\rm u} c_{\rm d} \sin\varphi \left
( s_{\rm u} s_{\rm d} c + c_{\rm u} c_{\rm d} \cos\varphi \right )}{s^2_{\rm
u} s^2_{\rm d} c^2 + c^2_{\rm u} c^2_{\rm d} + 2 s_{\rm u} c_{\rm u} s_{\rm d} c_{\rm d} c
\cos\varphi} \; , \nonumber \\ \nonumber \\
\sin (2\beta) & = & \frac{2 s_{\rm u} c_{\rm d} \sin\varphi \left
( c_{\rm u} s_{\rm d} c - s_{\rm u} c_{\rm d} \cos\varphi \right )}{c^2_{\rm
u} s^2_{\rm d} c^2 + s^2_{\rm u} c^2_{\rm d} - 2 s_{\rm u} c_{\rm u} s_{\rm d} c_{\rm d} c
\cos\varphi} \; .
\end{eqnarray}
To an excellent degree of accuracy, one finds $\alpha \approx
\varphi$. In order to illustrate how accurate this relation is, let us
input the central values of $\theta$, $\theta_{\rm u}$ and $\theta_{\rm 
d}$ (i.e., $\theta = 2.25^{\circ}$, $\theta_{\rm u} = 4.57^{\circ}$
and $\theta_{\rm d} = 12.7^{\circ}$) to Eq. (23). Then one arrives at
$\varphi - \alpha \approx 1^{\circ}$ as well as $\sin (2\alpha)
\approx 0.34$ and $\sin (2\beta) \approx 0.65$. 
It is expected that $\sin (2\alpha)$ and $\sin (2\beta)$
will be directly measured from the $CP$ asymmetries in 
$B_d \rightarrow \pi^+\pi^-$ and $B_d \rightarrow J /\psi K_S$ modes
at a $B$-meson factory.

Note that the three sides of the unitarity triangle 
(21) can be rescaled by $|V_{cb}|$. In a very good approximation
(with the relative error $\sim 2\%$), one arrives at
\begin{equation}
|S_u| ~ : ~ |S_c| ~ : ~ |S_t| \; \approx \; s_{\rm u} c_{\rm d} ~ : ~ 
s^{~}_{\rm C} ~ : ~ s_{\rm d} \; .
\end{equation}
Equivalently, one can obtain
\begin{equation}
s_{\alpha} ~ : ~ s^{~}_{\beta} ~ : ~ s_{\gamma} \; \approx \; s^{~}_{\rm C} 
~ : ~ s_{\rm u} c_{\rm d} ~ : ~ s_{\rm d} \; ,
\end{equation}
where $s_{\alpha} \equiv \sin\alpha$, etc.
The rescaled unitarity triangle is shown in Fig. 3(b). Comparing this
triangle with the Cabibbo triangle in Fig. 1, we find that they are 
indeed congruent with each other to a high degree of accuracy.
The congruent relation between these two triangles is particularly
interesting, since the Cabibbo triangle is essentially a feature of the physics
of the first two quark families, while the unitarity triangle is
linked to all three families. In this connection it is of special
interest to note that in models which specify the textures of the mass 
matrices the Cabibbo triangle and hence three inner angles of the unitarity
triangle can be fixed by the spectrum of the light quark masses and
the $CP$-violating phase $\varphi$ (see, e.g., Ref. \cite{FritzschXing95}).

k) It is worth pointing out that the u-channel and d-channel mixing
angles are related to the so-called Wolfenstein parameters 
\cite{Wolfenstein83} in a simple way:
\begin{eqnarray}
\tan\theta_{\rm u} & = & \left | \frac{V_{ub}}{V_{cb}} \right | 
\; \approx \; \lambda \sqrt{\rho^2 + \eta^2} \; , \; \nonumber \\
\tan\theta_{\rm d} & = & \left | \frac{V_{td}}{V_{ts}} \right |
\; \approx \; \lambda \sqrt{ (1-\rho)^2 + \eta^2} \; ,
\end{eqnarray}
where $\lambda \approx s^{~}_{\rm C}$ measures the magnitude of $V_{us}$.
Note that the $CP$-violating parameter $\eta$ is linked to $\varphi$
through
\begin{equation}
\sin\varphi \; \approx \; \frac{\eta}{\sqrt{\rho^2 + \eta^2}
\sqrt{(1-\rho)^2 + \eta^2}} \; 
\end{equation}
in the lowest-order approximation. Then $\varphi =90^{\circ}$ implies
$\eta^2 \approx \rho ( 1- \rho)$, on the condition $0 < \rho < 1$. In
this interesting case, of course, the flavor mixing matrix can 
fully be described in terms of only three independent parameters.

l) Compared with the standard parametrization of the flavor mixing
matrix $V$ advocated in
Ref. \cite{Standard}, the parametrization (8) has an additional
advantage: the renormalization-group evolution of $V$, from the weak
scale to an arbitrary high energy scale, is 
to a very good approximation associated only with the angle $\theta$. This can
easily be seen if one keeps the $t$ and $b$ Yukawa couplings only  
and neglects possible threshold effect in the one-loop
renormalization-group equations of the Yukawa matrices \cite{RGE}.
Thus the parameters $\theta_{\rm u}$, $\theta_{\rm d}$ and $\varphi$
are essentially independent of the energy scale, while $\theta$ does
depend on it and will change if the underlying scale is shifted, say
from the weak scale ($\sim 10^2$ GeV) to the grand unified theory
scale ($\sim 10^{16}$ GeV). In short, the heavy quark mixing is
subject to renormalization-group effects; but the u- and d-channel
mixings are not, likewise the phase $\varphi$ describing $CP$
violation.

In this paper we have presented a new description of the flavor mixing 
phenomenon, which is based on the phenomenological fact that the quark 
mass spectrum exhibits a clear hierarchy pattern. This leads uniquely
to the interpretation of the flavor mixing in terms of a heavy quark
mixing, followed by the u-channel and d-channel mixings. The complex
phase $\varphi$, describing the relative orientation of the u-channel
mixing and the d-channel mixing in the complex plane, signifies
$CP$ violation, which is a phenomenon primarily linked to the physics
of the first two families. The Cabibbo angle is not a basic mixing
parameter, but given by a superposition of two terms involving the
complex phase $\varphi$. The experimental data suggest that the phase
$\varphi$, which is directly linked to the phases of the quark mass
terms, is close to $90^{\circ}$. This opens the possibility to
interpret $CP$ violation as a maximal effect, in a similar way as
parity violation.

Our description of flavor mixing has many clear advantages compared
with other descriptions. We propose that it should be used in the
future description of flavor mixing and $CP$ violation, in particular, 
for the studies of quark mass matrices and $B$-meson physics.

{\large\it Acknowledgements:} ~ One of us (H.F.) likes to express his
gratitude to Prof. T. Maskawa and Prof. M. Ninomiya for their generous 
support during his stay at YITP in spring 1997.
His research was also supported in part by DFG-contract 412/22-1, EEC-contract 
SC1-CT91-0729 and EEC-contract CHRX-CT94-0579 (DG 12 COMA).
Z.Z.X. is grateful to Prof. A.I. Sanda for his warm hospitality
and to the Japan Society for the
Promotion of Science for its financial support.

\end{document}